\documentclass[11pt,a4paper]{article}
\usepackage{amsmath,amsfonts,amssymb,amsthm,amstext,amscd,array}
\usepackage[table]{xcolor}
\usepackage[pdftex]{graphicx}
\usepackage{tikz}
\usepackage{mathrsfs}
\usepackage{bm}
\usepackage[arrow,matrix,curve]{xy}
\usepackage{soul}
\usepackage{authblk}

\renewcommand{\thefootnote}{\fnsymbol{footnote}}
\newcommand{\newsection}{\setcounter{equation}{0}\section}
\renewcommand*{\thefootnote}{\fnsymbol{footnote}}

\usepackage{hyperref}
\hypersetup{colorlinks,%
citecolor=darkgray,%
filecolor=darkgray,%
linkcolor=darkgray,%
urlcolor=brown,%
unicode,pdffitwindow,pdftitle={},pdfauthor={}}

\begin{document}
\begin{titlepage}

\begin{center}

\vspace{3cm}

\baselineskip=24pt

{\LARGE\bf Minimal AdS-Lorentz supergravity in three-dimensions }

\baselineskip=14pt

\vspace{1cm}
{\bf O. Fierro}\footnote{Email: \ {\tt ofierro@ucsc.cl}},
{\bf F. Izaurieta}\footnote{Email: \ {\tt fizaurie@udec.cl}},
{\bf P. Salgado}\footnote{Email: \ {\tt pasalgad@udec.cl}} and
{\bf O. Valdivia}\footnote{Email: \ {\tt ovaldivi@unap.cl}}
\\[4mm]
\noindent $^{\S}${\it Facultad de Ingenier\'ia y Arquitectura,\\ Universidad Arturo Prat,\\Iquique, Chile}
\\[4mm]
\noindent $^{\dag}$$^{\ddagger}${\it Departamento de F\'isica,\\ Universidad de Concepci\'on, casilla 160-C,\\ Concepci\'on, Chile}
\\[4mm]
\noindent $^{\ast}${\it Departamento de Matem\'atica y F\'isica Aplicadas,\\ Universidad Cat\'olica de la Sant\'isima Concepci\'on,
Alonso de Rivera 2850,\\ Concepci\'on, Chile}
\\[30mm]

\end{center}

\begin{abstract}

\noindent The $\mathcal{N}=1$ AdS-Lorentz superalgebra is studied and its relationship to semigroup expansion developed. Using this mathematical tool, the invariant tensors and Casimir operators are found. In terms of these invariants, a three-dimensionnal Chern--Simons supergravity action with AdS-Lorentz symmetry is constructed. The Killing spinors for a BTZ black-hole like solution of the theory are discussed.

\baselineskip=12pt

\end{abstract}

\end{titlepage}
\setcounter{page}{2}

%\newpage

{\baselineskip=12pt
\tableofcontents
}

\renewcommand{\thefootnote}{\arabic{footnote}}
\setcounter{footnote}{0}

%\newpage

\newsection{Introduction}

Since the original works of Deser, Jackiw and 't Hooft~\cite{DESER1984220,DESER1984405}, three dimensional gravity has attracted attention. Despite having no propagating degrees of freedom, the BTZ black-hole solution~\cite{Banados:1992wn, Banados:1992gq} and the quantization of the theory by Witten~\cite{Wit88} are its highly non-trivial trademarks.
These features seem to be rooted in the fact that the Einstein--Hilbert (EH) Lagrangian with cosmological constant can be written (up to a boundary term) as a Chern--Simons (CS) three-form. Therefore, three-dimensional gravity corresponds to an off-shell quasi-invariant gauge theory (for AdS, dS or Poincar\'{e} depending on the cosmological constant).
The locally supersymmetric extension of Einstein gravity in three dimensions was carried out by Deser and Kay in Ref.~\cite{Deser:1982sw}. Regarding the CS formulation, three-dimensional supergravity arises very naturally in the case of negative~\cite{Achucarro:1987vz} and vanishing~\cite{Achucarro:1989gm,Howe:1995zm} cosmological constant. However, there is still the possibility of having other families of supergravity theories containing gauge groups larger than the supersymmetric AdS or Poincar\'e groups~\cite{Banh96a,Ede06b,Iza11a}. This is particularly interesting because symmetries enhancements usually invokes new generators in the Lie algebra. Subsequently, this requires the inclusion of extra gauge fields in the gauge potential, giving rise to non-minimal couplings of \textquotedblleft matter\textquotedblright\ fields with geometry in such a way that gauge invariance is preserved.

The purpose of this work is to analyse the construction of three-dimensional CS supergravity theories whose symmetry groups are obtained by an S-expansion of the $\mathcal{N}=1$ supersymmetric AdS algebra $\mathfrak{osp}(1|2)\otimes\mathfrak{sp}(2)$. The S-expansion method~\cite{Izaurieta:2006zz,Izaurieta:2009gc} is a powerful tool in order for obtaining new Lie algebras starting from a given one. Moreover, it provides the associated invariant tensors of the expanded algebra in a simple way. Since the invariant tensor is an essential ingredient in the construction of gauge theories and in particular of CS (super)gravities, it is a welcomed feature. 

The application of S-expansion methods in the context of supergravity was first introduced in~\cite{Edelstein:2006se} and subsequently in~\cite{Iza06c} as an attempt to describe the low energy regime of $M$-Theory. More recently, a wide range of theories of S-expanded (super)gravities have been studied in different contexts, and with different motivations (see for instance~\cite{Iza09b,Salgado:2014jka,Concha:2015woa,Concha:2016zdb} and references therein). Also, in Ref.~\cite{Diaz2012}, three-dimensional gravity is constructed using the semi-simple extension of the Poincar\'e algebra~\cite{Sor04,Soroka:2011tc} as a gauge symmetry. The Lie algebra behind this symmetry can be obtained as a S-expansion of AdS algebra $\mathfrak{so}(d-1,2)$.

%In this paper we follow the natural continuation of this line of reasoning: First, we obtain the supersymmetric extension for the semi-simple Poincar\'e algebra~\cite{Sor06,Sor10} in three-dimensions, via S-expansion of the AdS superalgebra $\mathfrak{osp}(1|2)\otimes\mathfrak{sp}(2)$. Second, we compute the corresponding Chern-Simons supergravity action and finally some aspects of the physical content of the resulting theory are pointed out.

This article is organized as follows: In section \ref{sect2}, the supersymmetric extension of the three-dimensional AdS-Lorentz algebra is written. %, and the relationship between two different presentations of the same superalgebra is clarified. %
 In section \ref{sec3}, we
review the general properties of the S-expansion method. Also, it is explicitly shown that three-dimensional AdS-Lorentz superalgebra corresponds to a S-expansion of the AdS superalgebra. The components of the invariant tensor are worked out. In section \ref{sec4} we extend the notion of S-expansion to Casimir operators and the invariant operators associated to the expanded superalgebra are constructed. Section \ref{sec5} is devoted to the analysis of three-dimensional AdS-Lorentz CS supergravity. Field equations and symmetry transformations are worked out. In section \ref{sec6} we compute stationary solutions and its Killing spinors equation are found. Finally, section \ref{sec7} concludes this paper with some remarks and future developments.

\newsection{AdS-Lorentz superalgebra}\label{sect2}

In~Ref.~\cite{Soroka:2011tc} the semi-simple extension of the Poincar\'e algebra $\mathfrak{iso}(d-1,1)$, generated by Lorentz rotations $\left\lbrace J_{ab}\right\rbrace$ and translations $\left\lbrace P_{a} \right\rbrace$, has been carried out by the inclusion of a second-rank tensor generator $\left\lbrace Z_{ab}\right\rbrace$. Interestingly, this Lie algebra enhancement is isomorphic to the direct sum of the AdS and Lorentz algebra  $\mathfrak{so}(d-1,2) \oplus \mathfrak{so}(d-1,1)$ in any dimension. More recently, it has been shown in Refs.~\cite{Diaz2012,Salgado:2014qqa} that the so called AdS-Lorentz algebra can be obtained as a S-expansion of the AdS algebra and its In\"{o}n\"{u}-Wigner contraction leads to the Maxwell algebra. %%%%%%%%%%%%%%~\cite{}. %%%%%%%%%%%%%%%%%%%%
The supersymmetric extension in four-dimensions has been also considered in~Refs.~\cite{Sor06,Sor10}. Remarkably, both algebras, pure bosonic and supersymmetric, are semi-simple in contrast to the (super) Poincar\'e algebras.

In this work we are interested in the $\mathcal{N}=1$ AdS-Lorentz superalgebra in three-dimensions, which is defined by the following commutation relations:
\begin{equation}%
\begin{array}
[c]{ll}%
\left[  J_{a},J_{b}\right]  =\epsilon_{abc}J^{c}\, , & \qquad \left[  Z_{a},Z_{b}\right]
=\epsilon_{abc}Z^{c}\, , \medskip \\
\left[  J_{a},P_{b}\right]  =\epsilon_{abc}P^{c}\, , &\qquad \left[  P_{a},P_{b}\right]
=\epsilon_{abc}Z^{c}\, , \medskip\\
\left[  J_{a},Z_{b}\right]  =\epsilon_{abc}Z^{c}\, , &\qquad \left[  Z_{a},P_{b}\right]
=\epsilon_{abc}P^{c}\, , \medskip\\
\left[  P_{a},Q_{\alpha}\right]  =-\frac{1}{2}\left(  \Gamma_{a}Q\right)
_{\alpha}\, ,&\qquad \left[  Z_{a},Q_{\alpha}\right]  =-\frac{1}{2}\left(  \Gamma
_{a}Q\right)  _{\alpha}\, , \medskip\\
\left[  J_{a},Q_{\alpha}\right]  =-\frac{1}{2}\left(  \Gamma_{a}Q\right)
_{\alpha}\, ,  &\qquad  \left\{  Q_{\alpha},Q_{\beta}\right\}  =\left(  \Gamma
_{a}\mathcal{C}\right)  _{\alpha\beta}\left(  P^{a}+Z^{a}\right)\, ,
\end{array}
\ \ \ \ \label{adsl1}%
\end{equation}
where $J_a$ denote the generators of the Lorentz subalgebra $\mathfrak{so}(2,1)$, $P_a$ the translations, $Z_{a}$ are a new set of non-abelian generators, and $Q_{\alpha}$ the supercharges. Lorentz indices $a,b,...=0,1,2$ are rised and lowered with the Minkowski metric $\eta_{ab}$, $\epsilon_{abc}$ is the three-dimensional Levi-Civita symbol. Greek indices $\alpha, \beta...=0,1$ are rised and lowered by  the charge conjugation matrix $\mathcal{C}$, and $\Gamma^{a}$ denote the $2\times2$ gamma matrices representation (see Appendix~\ref{Appen} for spinor conventions). 

In the following section, we show that the (super)AdS-Lorentz algebra (\ref{adsl1}) can be derived as an application of the S-expansion procedure.

\newsection{Abelian Semigroup Expansion}\label{sec3}
The Lie algebra expansion procedure was introduced for the first time in Ref.~\cite{Hatsuda:2001pp}, and subsequently studied in in Refs.~\cite{deAz02,deAz07}. In this expansion method, we must consider the % consists in considering the original algebra as described by its associated
Maurer-Cartan forms on the group manifold. Some of the group parameters are rescaled by a factor $\lambda$, and the Maurer-Cartan forms are expanded as a power series in $\lambda$. The series is finally truncated in such a way that the closure of the expanded algebra is assured.

In Refs.~\cite{Izaurieta:2006zz,Izaurieta:2009gc,Iza06a} a natural outgrowth of the power series expansion method was proposed. The idea is to start with a Lie algebra $\mathfrak{g}$ and to combine it with the binary product structure of an abelian semigroup $S$ in order to define a new Lie algebra. This new algebra is known in general as a S-expanded algebra. In fact, from \cite[Theorem 3.1]{Izaurieta:2006zz}, it is possible to prove that the direct product $S\times\mathfrak{g}$ retains Lie algebra structure (see also \cite{Hatsuda:2001pp,deAz02,deAz04}). The most relevant cases are provided when subalgebras of $S\times\mathfrak{g}$ can be systematically extracted. For instance, any Lie algebra can be written as a
direct sum of subspaces $\mathfrak{g}=\bigoplus\nolimits_{p\in I}V_{p}$, where $I$ is a set of indices. The subspace structure of the algebra can be analysed defining a mapping $i:I\times I\rightarrow 2^{I}$ such that the Lie algebra $\mathfrak{g}$ can be written as $\left[V_{p},V_{q}\right]  \subset\bigoplus\nolimits_{r\in i_{\left(p,q\right)  }}V_{r}$. Now, whenever the semigroup $S$ admits a decomposition $S=\bigcup\nolimits_{p\in I}S_{p}$, satisfying the \textit{resonant} condition $S_{p}$\textperiodcentered$S_{q}\subset\bigcap\nolimits_{r\in i_{\left(  p,q\right)  }}S_{r}$, then it follows that $\mathfrak{S}_{R}=\bigoplus\nolimits_{p\in I}S_{p}\times V_{p}$ is a subalgebra of $S\times\mathfrak{g}$~\cite[Theorem 4.2]{Izaurieta:2006zz}. The procedure is practical because the subspace structure is arbitrary, but we use it in order to codify our physicist's intuition on the meaning of the symmetry (e.g. a subspace corresponds to Lorentz transformations, another to AdS boosts, etc.). Thus, using the S-expansion it is possible to find bigger symmetries in a simple way, and to do this preserving some valuable structure from a physical point of view. Without it, constructing bigger symmetries requires long and careful work regarding the closure of Jacobi's identity (or the self-consistency of $\mathrm{d}^{2}=0$ when working with Maurer-Cartan forms).

The S-expansion procedure has already been used in different contexts with different motivations. For instance, the so called $\mathfrak{B}_{m}$-algebras~\cite{Iza09b} (also known as generalized Poincar\'{e} algebras), were constructed from the AdS-algebra and a particular semigroup\footnote{This semigroup is endowed with the multiplication rule $\lambda _{\alpha }$\textperiodcentered$\lambda _{\beta}=\lambda _{\alpha +\beta }$ when $\alpha +\beta \leq N+1;$ and $\lambda_{\alpha }$\textperiodcentered$\lambda _{\beta }=\lambda _{N+1}$ otherwise.} denoted by $S_{E}^{(N)}=\left\{ \lambda _{\alpha }\right\} _{\alpha =0}^{N+1}$.  Moreover, in Ref.~\cite{Salgado:2014qqa} the so-called  AdS-Lorentz algebra $\mathfrak{so}\left(d-1,1\right) \oplus \mathfrak{so}\left( d-1,2\right) $~\cite{Sor04,Soroka:2011tc,Sor06} is obtained by means of the S-expansion procedure with a semigroup\footnote{This semigroup is endowed with the multiplication rule $\lambda _{\alpha }$\textperiodcentered$\lambda _{\beta }=\lambda _{\alpha +\beta }$ when \ $\alpha+\beta \leq N;$ and $\lambda _{\alpha }$\textperiodcentered$\lambda _{\beta }=\lambda _{\alpha +\beta-2\left[ (N+1)/2\right] }$ otherwise.} denoted by $S_{\mathcal{M}}^{(N)}=\left\{ \lambda _{\alpha}\right\}_{\alpha=0}^{N}$. This later algebra is related to the so called Maxwell algebra \cite{Bacry:1970ye, Schrader:1972zd} via a contraction process \cite{Lukierski:2010dy}.

Another interesting application is in the context of non-relativistic algebras. Recently, in Ref. \cite{Gonzalez:2016xwo} it was shown that it is possible to obtain the non-relativistic versions of both generalized Poincar\'{e} algebras and generalized AdS-Lorentz algebras. These were called generalized Galilean type I and type II, denoted by $\mathrm{GB_{n}}$ and $\mathrm{GL_{n}}$ respectively. It seems likely that new non-relativistic CS gravity theories may be constructed following a similar procedure as the one presented in Ref.~\cite{Andringa:2010it}. Its symmetries would correspond to deformations of the symmetries of the Newton-Cartan formulation of Newtonian gravity. This problem will be addressed in the near future.

\subsection{S-expansion and the AdS superalgebra}

\begin{figure}[htb]
		\centering
		\includegraphics[scale=0.22]{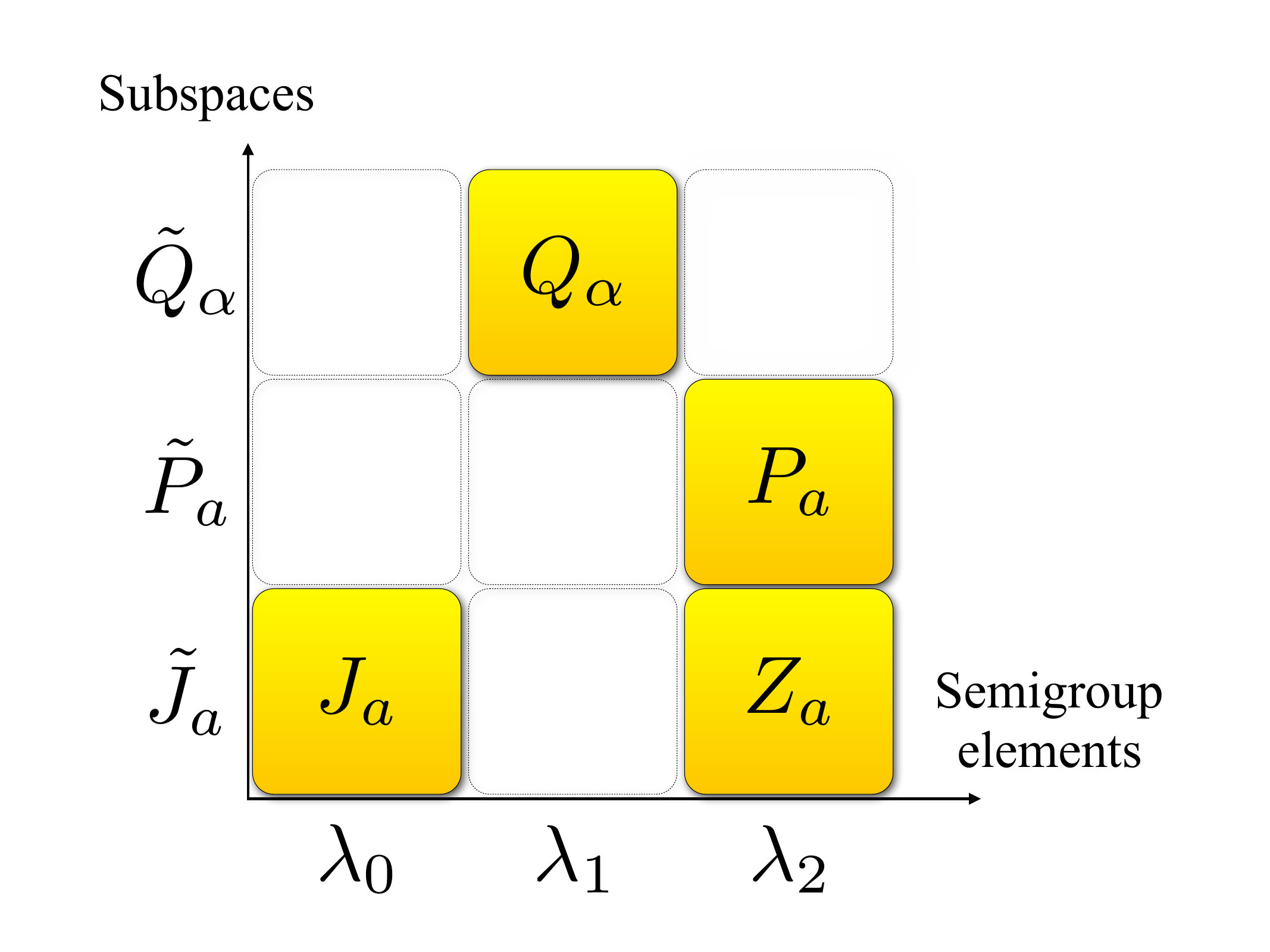}
		\caption{AdS-Lorentz superalgebra as resonant subalgebra of $S^{(2)}_{\mathcal{M}}\times\mathfrak{g}$}
		\label{fig3}
\end{figure}

In this section we construct the three-dimensional AdS-Lorentz superalgebra as a S-expansion of the AdS superalgebra $\mathfrak{g}=\mathfrak{osp}\left(\left.  2\right\vert 1\right)  \otimes\mathfrak{sp}\left(  2\right)  $, given by the commutation relations
\begin{equation}
\begin{array}
[c]{ll}%
\left[  \tilde{J}_{a},\tilde{J}_{b}\right]  =\epsilon_{abc}\tilde{J}^{c} \, , & \qquad
\left[  \tilde{P}_{a},\tilde{Q}_{\alpha}\right]  =-\frac{1}{2}\left(
\Gamma_{a}\tilde{Q}\right)  _{\alpha}\, , \medskip\\
\left[  \tilde{P}_{a},\tilde{P}_{b}\right]  =\epsilon_{abc}\tilde{J}^{c} \, ,  & \qquad
\left[  \tilde{J}_{a},\tilde{Q}_{\alpha}\right]  =-\frac{1}{2}\left(
\Gamma_{a}\tilde{Q}\right)  _{\alpha} \, , \medskip\\
\left[  \tilde{J}_{a},\tilde{P}_{b}\right]  =\epsilon_{abc}\tilde{P}^{c} \, , & \qquad
\left\{  Q_{\alpha},Q_{\beta}\right\}  =\left(  \Gamma_{a}\mathcal{C}\right)
_{\alpha\beta}\left(  \tilde{J}^{a}+\tilde{P}^{a}\right) \, .  %
\end{array} \label{sex1}
\end{equation}
Let us start by choosing the following subspace decomposition
\begin{equation}
\mathfrak{g}=V_{0}\oplus V_{1}\oplus V_{2}\, , \medskip \label{sex2}
\end{equation}
where $V_{0}=$Span$\left\{  \tilde{J}_{a}\right\}  $, $V_{1}=$Span$\left\{  \tilde{Q}_{\alpha}\right\}  $ and $V_{2}=$Span$\left\{  \tilde{P}_{a}\right\}  $. This decomposition obeys the following
structure%
\begin{equation}%
\begin{array}
[c]{lll}%
\left[  V_{0},V_{0}\right]  \subset V_{0}\, , & \qquad \left[  V_{0},V_{1}\right]
\subset V_{1}\, , & \qquad\left[  V_{0},V_{2}\right]  \subset V_{2}\, , \medskip\\
\left[  V_{1},V_{1}\right]  \subset V_{0}\oplus V_{2} \, , & \qquad \left[  V_{1}%
,V_{2}\right]  \subset V_{1}\, ,  &\qquad \left[  V_{2},V_{2}\right]  \subset V_{1} \, .%
\end{array}
\ \ \ \label{sex3}%
\end{equation}
At this point it is convenient to apply the S-expansion resonance theorem using~(\ref{sex3}) and a specific semigroup $S^{(2)}_{\mathcal{M}}$. A similar treatment was carried out in Ref.~\cite{Diaz2012} for the bosonic sector.
%\subsubsection{Semigroup $S^{(2)}_{\mathcal{M}}$}\label{Sec-SI}
Let $S^{(2)}_{\mathcal{M}}=\left\{  \lambda_{0},\lambda _{1},\lambda_{2}\right\}  $ be an abelian semigroup with multiplication law
\begin{equation}
\lambda_{\alpha}\text{\textperiodcentered}\lambda_{\beta}=\left\{
\begin{array}
[c]{c}%
\lambda_{\alpha+\beta},\quad\text{if }\alpha+\beta\leqslant2\\
\lambda_{\alpha+\beta-2}\quad\text{if }\alpha+\beta>2
\end{array}
\right.  \label{sex12}%
\end{equation}
or equivalently%
\begin{equation}
\begin{tabular}
[c]{l||lll}
& $\cellcolor{yellow!25}\lambda_{0}$ & $\cellcolor{blue!25}\lambda_{1}$ & $\cellcolor{purple!25}\lambda_{2}$\\\hline\hline
$\cellcolor{yellow!25}\lambda_{0}$ & $\cellcolor{yellow!25}\lambda_{0}$ & $\cellcolor{blue!25}\lambda_{1}$ & $\cellcolor{purple!25}\lambda_{2}$\\
$\cellcolor{blue!25}\lambda_{1}$ & $\cellcolor{blue!25}\lambda_{1}$ & $\cellcolor{purple!25}\lambda_{2}$ & $\cellcolor{blue!25}\lambda_{1}$\\
$\cellcolor{purple!25}\lambda_{2}$ & $\cellcolor{purple!25}\lambda_{2}$ & $\cellcolor{blue!25}\lambda_{1}$ & $\cellcolor{purple!25}\lambda_{2}$%
\end{tabular}
\label{sex5}%
\end{equation}
A particular partition for the semigroup $S^{(2)}_{\mathcal{M}}$ is given by
\begin{align}
S^{(2)}_{\mathcal{M}}  &  =S_{0}\cup S_{1}\cup S_{2}\, , \medskip\nonumber\\
&  =\left\{  \lambda_{0},\lambda_{2}\right\}  \cup\left\{  \lambda
_{1}\right\}  \cup\left\{  \lambda_{2}\right\} \, , \label{sex4}%
\end{align}
where the subsets $\lbrace S_{i}\rbrace_{i=0,1,2}$ obey
\begin{equation}
\begin{array}
[c]{lll}%
S_{0} $\textperiodcentered$ S_{0}\subset S_{0}\, , &\qquad S_{0} $\textperiodcentered$
S_{1}\subset S_{1} \, , &\qquad S_{0}$\textperiodcentered$ S_{2}\subset S_{2} \, , \medskip\\
S_{1}$\textperiodcentered$ S_{1}\subset S_{0}\cap S_{2}\, , & \qquad S_{1}%
$\textperiodcentered$ S_{2}\subset S_{1}\, , & \qquad S_{2}$\textperiodcentered$
S_{2}\subset S_{0}\, .
\end{array}
\ \ \label{sex6}%
\end{equation}
Comparing (\ref{sex6}) with (\ref{sex3}), one finds that the resonant condition~\cite[Theorem 4.2]{Izaurieta:2006zz} is satisfied. Therefore, a subalgebra 
\begin{equation}
\mathfrak{G}_{R}=W_{0}\oplus W_{1}\oplus W_{2}\, , \label{sex7}
\end{equation}
can be extracted with
\begin{align}
W_{0}  &  \equiv S_{0}\times V_{0}=\mathrm{Span}\, \left\{  \lambda_{0}\tilde{J}_{a},\lambda_{2}\tilde{J}_{a}\right\}\, ,
\medskip\label{sex8}\\
W_{1}  &  \equiv S_{1}\times V_{1}=\mathrm{Span}\, \left\{  \lambda_{1}\tilde{Q}_{\alpha}\right\} \, , \medskip\label{sex9}\\
W_{2}  &  \equiv S_{2}\times V_{2}=\mathrm{Span}\, \left\{  \lambda_{2}\tilde{P}_{a}\right\} \, , \label{sex10}
\end{align}
which corresponds in this case to the minimal AdS-Lorentz Lie superalgebra. In fact, computing commutation relations 
\begin{equation}%
\begin{tabular}
[c]{ll}%
$\left[  \lambda_{0}\tilde{J}_{a},\lambda_{0}\tilde{J}_{b}\right]
=\lambda_{0}\left[  \tilde{J}_{a},\tilde{J}_{b}\right]  =\lambda_{0}%
\epsilon_{abc}\tilde{J}^{c}$\,, & \quad\medskip$\left[  \lambda_{2}\tilde{P}_{a},\lambda
_{2}\tilde{P}_{b}\right]  =\lambda_{2}\left[  \tilde{P}_{a},\tilde{P}%
_{b}\right]  =\lambda_{2}\epsilon_{abc}\tilde{P}^{c}$\,,\\
$\left[  \lambda_{0}\tilde{J}_{a},\lambda_{2}\tilde{J}_{b}\right]
=\lambda_{2}\left[  \tilde{J}_{a},\tilde{J}_{b}\right]  =\lambda_{2}%
\epsilon_{abc}\tilde{J}^{c}$\,, & \quad\medskip$\left[  \lambda_{0}\tilde{J}_{a},\lambda
_{1}\tilde{Q}_{\alpha}\right]  =\lambda_{1}\left[  \tilde{J}_{a},\tilde
{Q}_{\alpha}\right]  =-\frac{\lambda_{1}}{2}\left(  \Gamma_{a}\tilde
{Q}\right)  _{\alpha}$\,,\\
$\left[  \lambda_{2}\tilde{J}_{a},\lambda_{2}\tilde{J}_{b}\right]
=\lambda_{2}\left[  \tilde{J}_{a},\tilde{J}_{b}\right]  =\lambda_{2}%
\epsilon_{abc}\tilde{J}^{c}$\,, &\quad\medskip $\left[  \lambda_{2}\tilde{J}_{a},\lambda
_{1}\tilde{Q}_{\alpha}\right]  =\lambda_{1}\left[  \tilde{J}_{a},\tilde
{Q}_{\alpha}\right]  =-\frac{\lambda_{1}}{2}\left(  \Gamma_{a}\tilde
{Q}\right)  _{\alpha}$\,,\\
$\left[  \lambda_{0}\tilde{J}_{a},\lambda_{2}\tilde{P}_{b}\right]
=\lambda_{2}\left[  \tilde{J}_{a},\tilde{P}_{b}\right]  =\lambda_{2}%
\epsilon_{abc}\tilde{P}^{c}$\,, &\quad\medskip $\left[  \lambda_{2}\tilde{P}_{a},\lambda
_{1}\tilde{Q}_{\alpha}\right]  =\lambda_{1}\left[  \tilde{P}_{a},\tilde
{Q}_{\alpha}\right]  =-\frac{\lambda_{1}}{2}\left(  \Gamma_{a}\tilde
{Q}\right)  _{\alpha}$\,,\\
$\left[  \lambda_{2}\tilde{J}_{a},\lambda_{2}\tilde{P}_{b}\right]
=\lambda_{2}\left[  \tilde{J}_{a},\tilde{P}_{b}\right]  =\lambda_{2}%
\epsilon_{abc}\tilde{P}^{c}$\,, &\quad\medskip $\left\{  \lambda_{1}\tilde{Q}_{\alpha}%
,\lambda_{1}\tilde{Q}_{\beta}\right\}  =\lambda_{2}\left\{  \tilde{Q}_{\alpha
},\tilde{Q}_{\beta}\right\}  =\lambda_{2}\left(  \Gamma_{a}\mathcal{C}\right)
_{\alpha\beta}\left(  \tilde{J}_{a}+\tilde{P}_{a}\right)  $\,,%
\end{tabular}
\label{xaxa}%
\end{equation}
and renaming generators according to
\begin{equation}
\begin{array}
[c]{ll}%
J_{a}\equiv \lambda_{0}\tilde{J}_{a}\, , & \qquad Z_{a}\equiv \lambda_{2}\tilde{J}_{a}\, ,\medskip\\
P_{a}\equiv \lambda_{2}\tilde{P}_{a} \, ,& \qquad Q_{\alpha}\equiv \lambda_{1}\tilde{Q}_{\alpha} \, ,%
\end{array}
\ \ \label{sex11}
\end{equation}
it is straightforward to check that $\mathfrak{G}_{R}=W_{0}\oplus W_{1}\oplus W_{2}$ corresponds to  (\ref{adsl1}). 
This way, the AdS-Lorentz superalgebra is a resonant subalgebra of $S^{(2)}_{\mathcal{M}}\times\mathfrak{g}$. In order to get a better intuition about the S-expansion and resonant subalgebra procedure, it is helpful to use a diagram as shown in figure~(\ref{fig3}).
Subspaces of $\mathfrak{g}$ are represented in vertical axis while $S^{(2)}_{\mathcal{M}}$ elements are placed in horizontal axis. Coloured regions corresponds to the resonant subalgebra $\mathfrak{G}$ with respect to $S^{(2)}_{\mathcal{M}}\times\mathfrak{g}$.  

\subsection{Invariant Tensors}

The problem of finding all the invariant tensors associated to a given Lie algebra is, to the best of our knowledge, not completely understood. From the physical point of view, this limitation has direct impact on the construction of topological theories of gravity such as CS. However, as we will see below, there are complementary S-expansion theorems which allows us to find the invariant tensors and consequently the Casimir operators. These are easily constructed in terms of the invariants of the original algebra and the semigroup structure.

For the current case, recall the invariants of the AdS superalgebra $\mathfrak{g}=\mathfrak{osp}\left(  \left.  2\right\vert 1\right)  \otimes\mathfrak{sp}\left(  2\right)\,$,
\begin{equation}%
\begin{array}
[c]{lll}%
\left\langle \tilde{J}_{a}\tilde{J}_{b}\right\rangle =\tilde{\mu}\eta_{ab}\, , &\qquad
\left\langle \tilde{J}_{a}\tilde{P}_{b}\right\rangle =\tilde{\nu}\eta_{ab}\, , & \qquad
\left\langle \tilde{P}_{a}\tilde{P}_{b}\right\rangle =\tilde{\mu}\eta_{ab}\, , \medskip\\
\left\langle \tilde{P}_{a}\tilde{Q}_{\alpha}\right\rangle =0\, , &\qquad \left\langle
\tilde{J}_{a}\tilde{Q}_{\alpha}\right\rangle =0\, , &\qquad \left\langle \tilde
{Q}_{\alpha}\tilde{Q}_{\beta}\right\rangle =2\left(  \tilde{\nu}-\tilde{\mu
}\right)  \mathcal{C}_{\alpha\beta}\, ,%
\end{array}
\label{it2}%
\end{equation}
where $\tilde{\mu}$ and $\tilde{\nu}$ are arbitrary real constants. Following~\cite[Theorem 7.1 and 7.2]{Izaurieta:2006zz} it is direct to show that the invariant tensor associated to the AdS-Lorentz superalgebra has the following components
\begin{equation}%
\begin{tabular}
[c]{lll}%
$\left\langle J_{a}J_{b}\right\rangle =\mu_{1}\eta_{ab}$\,,&  &\qquad $\left\langle
P_{a}P_{b}\right\rangle =\mu_{0}\eta_{ab}$\,, \medskip\\
$\left\langle J_{a}Z_{b}\right\rangle =\mu_{0}\eta_{ab}$\,,&  &\qquad $\left\langle
Q_{\alpha}Q_{\beta}\right\rangle =2\left(  \nu_{0}-\mu_{0}\right)
\mathcal{C}_{\alpha\beta}$\,,\medskip\\
$\left\langle J_{a}P_{b}\right\rangle =\nu_{0}\eta_{ab}$\,, &  &\qquad $\left\langle
J_{a}Q_{\alpha}\right\rangle =0$\,,\medskip\\
$\left\langle Z_{a}Z_{b}\right\rangle =\mu_{0}\eta_{ab}$\,, &  &\qquad $\left\langle
Z_{a}Q_{\alpha}\right\rangle =0$\,,\medskip\\
$\left\langle Z_{a}P_{b}\right\rangle =\nu_{0}\eta_{ab}$\,,&  & \qquad$\left\langle
P_{a}Q_{\alpha}\right\rangle =0$\,,%
\end{tabular}
\ \ \label{it3}%
\end{equation}
where
\begin{equation}
\mu_{1}\equiv\alpha_{1}\tilde{\mu}\, ,\qquad\mu_{0}\equiv\alpha_{0}\tilde{\mu
}\, ,\qquad\nu_{0}\equiv\alpha_{0}\tilde{\nu}\, , \label{it4}%
\end{equation}
are redefinitions for arbitrary real constants\footnote{$\alpha_0$ and $\alpha_1$ appear through \cite[Theorem 7.1]{Izaurieta:2006zz} }. In what follows, we make use of (\ref{it3}) for the construction of the S-expanded Casimir operators. We have to notice that one of the three constants in (\ref{it3}) can always be reabsorbed in a global multiplicative constant. Therefore, we can expect only two independent Casimir operators.

\newsection{Casimir Operators and the AdS-Lorentz superalgebra}\label{sec4}

In order to find the Casimir operators for the AdS-Lorentz superalgebra one needs to specify the Casimir operators of the original $\mathfrak{g}=\mathfrak{osp}\left(  2\left\vert 1\right. \right)  \otimes\mathfrak{sp}\left(  2\right)  $ superalgebra. Since $\mathfrak{g}$ is semi-simple, it has a nondegenerate Killing metric
\begin{equation}
k_{AB}=\mathrm{STr}\left(  T_{A}T_{B}\right)  =\left\langle
T_{A}T_{B}\right\rangle
\label{co1}
\end{equation}
which can be read from (\ref{it2}). Here, $\lbrace T_{A}\rbrace$ with $A=1,\ldots,\mathrm{dim}\mathfrak{g}$, denote the generators of $\mathfrak{g}$. In this way, assuming nonvanishing and $\tilde{\mu}\neq\tilde{\nu}$ real constants, the inverse components\footnote{Doted latin indices are running along the AdS-Boosts $\left\{\tilde{P}_{a}\right\}  $.} of  $k^{AB}$ are the following:
\begin{equation}
\begin{tabular}
[c]{lll}%
$k^{ab}=\frac{\tilde{\mu}}{\left(  \tilde{\mu}^{2}-\tilde{\nu}^{2}\right)  }%
\eta^{ab}$\,, &\qquad $k^{a\dot{b}}=-\frac{\tilde{\nu}}{\left(  \tilde{\mu}^{2}-\tilde{\nu
	}^{2}\right)  }\eta^{ab}$\,, &\qquad $k^{\dot{a}\dot{b}}=\frac{\tilde{\mu}}{\left(
	\tilde{\mu}^{2}-\tilde{\nu}^{2}\right)  }\eta^{\dot{a}\dot{b}}$\,, \medskip\\
$k^{a\alpha}=0$\,, &\qquad $k^{\dot{a}\alpha}=0$\,, &\qquad $k^{\alpha\beta}=-\frac{1}{2\left(
	\tilde{\mu}-\tilde{\nu}\right)  }\mathcal{C}^{\alpha\beta}$\,.%
\end{tabular}
\label{co4}
\end{equation}
We look for Casimir operators of degree two $C=C^{AB}T_{A}T_{B}$, where $C^{AB}$ are given by the components of the symmetric invariant tensor $\left\langle T_{A}T_{B}\right\rangle$. Using (\ref{it2}) and the generators of $\mathfrak{g}$, direct calculation shows
\begin{align}
C&=k^{AB}{T}_{A}{T}_{B}\, ,\label{co5}\\
 &=\frac{1}{\left(  \tilde{\mu}^{2}-\tilde{\nu}^{2}\right)  }\left[  \tilde{\mu}\left(  \tilde{J}^{a}\tilde{J}_{a}+\tilde{P}^{a}\tilde{P}_{a}+\frac{1}{2}\tilde{Q}^{\alpha}\tilde{Q}_{\alpha}\right)  +\tilde{\nu}\left( - \tilde{P}^{a}\tilde{J}{a}-\tilde{J}^{a}\tilde{P}_{a}+\frac{1}{2}\tilde{Q}^{\alpha}\tilde{Q}_{\alpha}\right)  \right] \,.
\label{co6}
\end{align}
From eq.(\ref{co6}), one clearly sees that the AdS superalgebra has two independent Casimir operators 
\begin{align}
C_{1}  & =\tilde{J}^{a}\tilde{J}_{a}+\tilde{P}^{a}\tilde{P}_{a}+\frac{1}{2}\tilde
{Q}^{\alpha}\tilde{Q}_{\alpha}\, ,\medskip \nonumber\\
C_{2}  & =-\tilde{P}^{a}\tilde{J}_{a}-\tilde{J}^{a}\tilde{P}_{a}+\frac{1}{2}\tilde
{Q}^{\alpha}\tilde{Q}_{\alpha}\, .%
\label{co7}
\end{align}

\subsection{AdS-Lorentz Casimir operators}

Following Ref.~\cite{Diaz2012}, the Casimir operator for a S-expanded Lie algebra is defined by
\begin{equation}
C_{\mathrm{S-\exp}}=m^{\alpha\beta}C^{AB}T_{\left(  A,\alpha\right)  }T_{\left(
	B,\beta\right)  }\, ,%
\label{co8}
\end{equation}
where        $T_{\left(A,\alpha\right)}=\lambda_{\alpha}T_{A}$ denote the expanded generators and $m^{\alpha\beta}$ is the inverse of the
matrix $m_{\alpha\beta}=\alpha_{\gamma}K_{\alpha\beta}{}^{\gamma}$. Here $\alpha_{\gamma}$ denote arbitrary constants and $K_{\alpha\beta}{}^{\gamma}$ codifies the semigroup product law trough the definition
\begin{equation}
 K_{\alpha \beta}{}^{\gamma}=\left \{
 \begin{array}
[c]{l}
1\text{, when }\lambda_{\alpha}\lambda_{\beta}=\lambda_{\gamma}\text{,}\\
0\text{, otherwise.}%
\end{array}
\right.
\end{equation}

In the case of $S^{(2)}_{\mathcal{M}}$, the  $K_{\alpha \beta}{}^{\gamma}$ are given by
\begin{equation}
\begin{tabular}
[c]{lll}%
$K_{\alpha\beta}^{~~0}=%
\begin{pmatrix}
1 & 0 & 0\\
0 & 0 & 0\\
0 & 0 & 0
\end{pmatrix}
$\,, &\qquad $K_{\alpha\beta}^{~~1}=%
\begin{pmatrix}
0 & 1 & 0\\
1 & 0 & 1\\
0 & 1 & 0
\end{pmatrix}
$\,, &\qquad $K_{\alpha\beta}^{~~2}=%
\begin{pmatrix}
0 & 0 & 1\\
0 & 1 & 0\\
1 & 0 & 1
\end{pmatrix}
$\,.%
\end{tabular}
\ \ \
\label{co9}
\end{equation}
Therefore, the metric $m_{\alpha\beta}$ for $S^{(2)}_{\mathcal{M}}$ corresponds to
\begin{equation}
m_{\alpha\beta}=\alpha_{\gamma}K_{\alpha\beta}^{~~\gamma}=
\begin{pmatrix}
\alpha_{0} & \alpha_{1} & \alpha_{2}\\
\alpha_{1} & \alpha_{2} & \alpha_{1}\\
\alpha_{2} & \alpha_{1} & \alpha_{2}%
\end{pmatrix}\, ,
\label{co10}
\end{equation}
and its inverse $m^{\alpha\beta}$ reads%
\begin{equation}
m^{\alpha\beta}=\frac{1}{\det m_{\alpha\beta}}%
\begin{pmatrix}
\alpha_{2}^{2}-\alpha_{1}^{2} & 0 & -\left(  \alpha_{2}^{2}-\alpha_{1}%
^{2}\right)  \\
0 & \alpha_{2}\left(  \alpha_{0}-\alpha_{2}\right)   & -\alpha_{1}\left(
\alpha_{0}-\alpha_{2}\right)  \\
-\left(  \alpha_{2}^{2}-\alpha_{1}^{2}\right)   & -\alpha_{1}\left(
\alpha_{0}-\alpha_{2}\right)   & \alpha_{0}\alpha_{2}-\alpha_{1}^{2}%
\end{pmatrix}\, ,
\label{co11}
\end{equation}
where the constants $\alpha_{0}$, $\alpha_{1}$ and $\alpha_{2}$ must satisfy
\begin{equation}
\det m_{\alpha\beta}=\left(  \alpha_{0}-\alpha_{2}\right)  \left(  \alpha
_{2}^{2}-\alpha_{1}^{2}\right)  \neq0 \, .
\label{co12}
\end{equation}
Using (\ref{co7}), (\ref{co11}) in (\ref{co8}) and defining
\begin{align*}
\alpha & \equiv\alpha_{2}\alpha_{0}-\alpha_{2}^{2}\, , \medskip\\
\beta & \equiv\alpha_{2}\alpha_{0}-\alpha_{1}^{2}\, ,%
\end{align*}
we get two independent S-expanded Casimir operators for the AdS-Lorentz superalgebra,

\begin{align}
C_{\mathrm{S-\exp1}}  & =\frac{1}{2}\bar{Q}Q-J^{2}+2\left(  J^{a}Z_{a}-J^{a}P_{a}\right)\, ,\nonumber \medskip   \\
C_{\mathrm{S-\exp2}}  & =J^{2}+Z^{2}+P^{2}-2\left(  J^{a}Z_{a}-J^{a}P_{a}+Z^{a}P_{a}\right)\, . \label{sexco}
\end{align}

\newsection{Chern--Simons supergravity}\label{sec5}

We are interested in a supergravity theory which is invariant under an expanded AdS symmetry, such that it contains the EH term, the exotic gravitational CS term, plus cosmological constant as a certain limit. A type of symmetry fulfilling these conditions is precisely the minimal AdS-Lorentz supralgebra.

The fundamental field we consider is the one-form gauge potential $A=A^{a}_{\mu}dx^{\mu}\otimes T_{a}\,$, taking values in the Lie algebra (\ref{adsl1})
\begin{equation}
A=\frac{1}{\ell}e^{a}P_{a}+\omega^{a}J_{a}+\sigma^{a}Z_{a}+\frac{1}{\sqrt{\ell}}\bar{\psi}Q\, . \label{cs1}
\end{equation}
Here $e^{a}(x)$ is the vielbein, $\ell$ is a constant length parameter, $\omega^{a}(x)$ the spin connection and $\psi^{\alpha}(x)$ is a spin 3/2 gravitino. Moreover, the one-form $\sigma^{a}(x)$ will be referred to as the Lorentz gauge field since, as it will be shown later, it transforms as a vector under local Lorentz transformations.

In principle, it suffices to use the connection eq.(\ref{cs1}) in the canonical CS action functional~\cite{Zanelli:2005sa}
\begin{equation}
S_{\mathrm{cs}}^{\left(  2+1\right)  }\left[  A\right]  =\frac{\kappa}{4\pi}\int_{M_{3}}\left\langle A\wedge\left(  \mathrm{d}A+\frac{2}{3}A\wedge A\right)  \right\rangle  \label{cs7}
\end{equation}
in order to have the AdS-Lorentz supergravity theory. However, in order to gain some physical intuition on the Lagrangian terms it is convenient to use the subspace separation method (SSM). The applicability of the method relies on considering the CS $\left(2n+1\right)$-form as a particular case of a transgression form~\cite{Mor06a}. Following Ref.~\cite{Iza06a}, we write the triangle equation

\begin{equation}
\mathcal{Q}_{A_{2}\leftarrow A_{0}}^{\left(  2n+1\right)  }=\mathcal{Q}_{A_{2}\leftarrow A_{1}%
}^{\left(  2n+1\right)  }+\mathcal{Q}_{A_{1}\leftarrow A_{0}}^{\left(  2n+1\right)
}+\mathrm{d}\mathcal{Q}_{A_{2}\leftarrow A_{1}\leftarrow A_{0}}^{\left(  2n\right)  } \, ,\label{cs8}%
\end{equation}
which decompose the transgression form as a sum of two transgressions depending on
an intermediate connection, where each transgression is defined by
\begin{equation}
\mathcal{Q}_{A\leftarrow\bar{A}}^{\left(  2n+1\right)  }=\left(  n+1\right)  \int
_{0}^{1}dt\left\langle \left(  A-\bar{A}\right)\wedge F_{t}^{n}\right\rangle
\label{cs15}%
\end{equation}
with $F_{t}=\mathrm{d}A_{t}+A_{t}\wedge A_{t}$ and $A_{t}=\left(  A-\bar{A}\right)  t+\bar
{A}$. The last term in eq.(\ref{cs8}) is given by
\begin{equation}
\mathcal{Q}_{A\leftarrow\bar{A}\leftarrow\tilde{A}}^{\left(  2n\right)  }=n\left(
n+1\right)  \int_{0}^{1}dt\int_{0}^{t}ds\left\langle \left(  A-\bar{A}\right)\wedge
\left(  \bar{A}-\tilde{A}\right)  \wedge F_{st}^{n-1}\right\rangle \label{cs16}%
\end{equation}
where $F_{st}=\mathrm{d}A_{s,t}+A_{s,t}\wedge A_{s,t}$ and $A_{s,t}=\left(  A-\bar{A}\right)
s+\left(  \bar{A}-\tilde{A}\right)  t+\tilde{A}$ (see Ref.~\cite[Chapter 2 and 3]{Valdivia:2014zla} for further details).

The SSM embodies the following steps:
\begin{enumerate}
\item Split the superalgebra into $p+1$ subspaces $\mathfrak{g}=V_{0}%
\oplus...\oplus V_{p}$ \, .

\item Write the connection as a sum of pieces valued on every subspace
$A=a_{0}+...+a_{p}$, $\bar{A}=\bar{a}_{0}+...+\bar{a}_{p}$ with $a_{i},\bar
{a}_{i}\in\mathfrak{g}$ for $i=0,1,...,p$ \, .

\item Evaluate the triangle equation (\ref{cs8}) with the connections written
in terms of pieces valued in every subspace%
\begin{equation}
\begin{tabular}
[c]{lll}%
$A_{0}=\bar{A}$\, , &\qquad $A_{1}=a_{0}+...+a_{p-1}\, ,$ &\qquad $A_{2}=A\, .$%
\end{tabular}
\end{equation}

\item Repeat step $3$ for the transgression $\mathcal{Q}_{A_{1}\leftarrow A_{0}}^{\left(
2n+1\right)  }$ and so on.
\end{enumerate}

For the present case, $n=1$ and the AdS-Lorentz algebra splits into subspaces $\mathfrak{g}=V_{0}\oplus V_{1}\oplus V_{2}\oplus V_{3}$
with%
\begin{equation}%
\begin{tabular}
[c]{llll}%
$V_{0}=$Span$
\left\{  J_{a}\right\}\, , $ &\quad $V_{1}=$Span$\left\{  Z_{a}\right\}\, ,  $ &\quad $V_{2}=$Span$\left\{  P_{a}\right\}\, ,  $ &\quad $V_{3}=$Span$\left\{  Q_{\alpha}\right\}\, .  $%
\end{tabular}
\ \label{cs9}%
\end{equation}
Using the intermediate connections
\begin{align}
A_{0}  &  =0 \, ,\label{cs2}\\
A_{1}  &  =\omega \, ,\label{cs3}\\
A_{2}  &  =\sigma+\omega\, ,\label{cs4}\\
A_{3}  &  =e+\sigma+\omega\, ,\label{cs5}\\
A_{4}  &  =\bar{\psi}+e+\sigma+\omega\, ,\label{cs6}%
\end{align}
with%
\begin{equation}%
\begin{tabular}
[c]{llll}%
$\omega=\omega^{a}J_{a}\, ,$ &\quad $\sigma=\sigma^{a}Z_{a}\, ,$ &\quad $e=\frac{1}{\ell}%
e^{a}P_{a}\, ,$ & \quad $\bar{\psi}=\frac{1}{\sqrt{\ell}}\psi^{\alpha}Q_{\alpha}\, ,$
\end{tabular}
\ \label{cs11}%
\end{equation}
and applying (\ref{cs8}) recursively, we find\footnote{In what follows we omit the wedge symbol \textquotedblleft$\wedge$\textquotedblright\ in order to have shorter expressions.}
\begin{align}
&  \mathcal{Q}_{A_{4}\leftarrow A_{3}}^{\left(  3\right)  }=\frac{1}{\ell}\left(
\nu_{0}-\mu_{0}\right)  \left[  \left( \frac{1}{\ell} e^{a}+\sigma^{a}\right)
\left(  \bar{\psi}\Gamma_{a}\psi\right)  -2\bar{\psi}\mathrm{D}\psi\right] \, ,
\medskip\label{cs12}\\
&  \mathcal{Q}_{A_{3}\leftarrow A_{2}}^{\left(  3\right)  }=\frac{\mu_{0}}{\ell^{2}%
}\left[  e_{a}T^{a}+\epsilon_{abc}e^{a}\sigma^{b}e^{c}\right]
+\frac{\nu_{0}}{\ell}\left[  \epsilon_{abc}e^{a}\left(
\sigma^{b}\sigma^{c}+\frac{1}{3\ell^{2}}e^{b}e^{c}\right)  +2e_{a}\left(  R^{a}%
+\mathrm{D}\sigma^{a}\right)  \right] \, , \medskip \label{cs13}\\
&  \mathcal{Q}_{A_{2}\leftarrow A_{1}}^{\left(  3\right)  }=\mu_{0}\left[
\sigma_{a}\left(  2R^{a}+\mathrm{D}\sigma^{a}\right)  +\frac{1}{3}\epsilon_{abc}\sigma^{a}\sigma^{b}\sigma^{c}\right] \, , \medskip \label{cs14}\\
&  \mathcal{Q}_{A_{1}\leftarrow A_{0}}^{\left(  3\right)  }=\mu_{1}\left[  \omega
^{a}\mathrm{d}\omega_{a}+\frac{1}{3}\epsilon_{abc}\omega^{a}\omega^{b}\omega
^{c}\right]  \, , \medskip\label{cs20}\\
&  \mathcal{Q}_{A_{4}\leftarrow A_{3}\leftarrow A_{0}}^{\left(  2\right)  }%
=0\, , \medskip\label{cs21}\\
&  \mathcal{Q}_{A_{3}\leftarrow A_{2}\leftarrow A_{0}}^{\left(  2\right)  }=\frac
{\nu_{0}}{\ell}e^{a}\left(  \omega_{a}+\sigma_{a}\right) \, , \medskip
\label{cs22}\\
&  \mathcal{Q}_{A_{2}\leftarrow A_{1}\leftarrow A_{0}}^{\left(  2\right)  }=\mu_{0}\sigma^{a}\omega_{a} \, ,\label{cs23}%
\end{align}
where%
\begin{equation}
\begin{array}[c]{ll}
	T^{a}   = \mathrm{d}e^{a}+\epsilon_{~bc}^{a}\omega^{b}e^{c}\, , &\qquad \mathrm{D}\sigma^{a}    =\mathrm{d}\sigma^{a}+\epsilon_{~bc}^{a}\omega^{b}\sigma^{c} \, , \medskip\\
	R^{a}   =\mathrm{d}\omega^{a}+\frac{1}{2}\epsilon_{~bc}^{a}\omega^{b}\omega
	^{c} \, ,
	& \qquad \mathrm{D}\bar{\psi}   =\mathrm{d}\bar{\psi}-\frac{1}{2}\omega^{a}\left(  \bar{\psi}\Gamma
	_{a}\right)\, .
	\end{array}
	\label{cs24}
	\end{equation}
	
Since $\mathcal{Q}_{A_{4}\leftarrow A_{0}}^{\left(  3\right)  }=
\mathcal{L}_{\mathrm{cs}}^{\left(  3\right)  }(A)$, the CS Lagrangian can be read by
collecting (\ref{cs12})-(\ref{cs23})%
\begin{align}
\mathcal{L}_{\mathrm{cs}}^{\left(  3\right)  }  &  =\kappa\frac{\nu_{0}}{\ell}\left[  \epsilon_{abc}e^{a}\left(  \sigma^{b}%
\sigma^{c}+\frac{1}{3\ell^{2}}e^{b}e^{c}\right)  +2e_{a}\left(  R^{a}+\mathrm{D}\sigma^{a}\right)  +\left(  \frac{1}{\ell}e^{a}+\sigma^{a}\right)  \left(
\bar{\psi}\Gamma_{a}\psi\right)  -2\bar{\psi}\mathrm{D}\psi\right] \medskip \nonumber\\
&  +\kappa\mu_{0}\left[  \sigma_{a}\left(  2R^{a}+%
\mathrm{D}\sigma^{a}\right)  +\epsilon_{abc}\sigma^{a}\left(
\frac{1}{\ell^{2}}e^{b}e^{c}+\frac{1}{3}\sigma^{b}\sigma^{c}\right)  +\frac{1}{\ell^{2}}e_{a}%
T^{a}-\frac{1}{\ell^{2}}\left(  e^{a}+\sigma^{a}\right)  \left(  \bar{\psi}%
\Gamma_{a}\psi\right)  +\frac{2}{\ell}\bar{\psi}\mathrm{D}\psi\right] \medskip\nonumber\\
&  +\kappa\mu_{1}\left[  \omega^{a}d\omega_{a}+\frac{1}{3}\epsilon_{abc}\omega
^{a}\omega^{b}\omega^{c}\right]  +\mathrm{d} \left[  \frac{\nu_{0}}{\ell}%
e^{a}\left(  \omega_{a}+\sigma_{a}\right)  +\mu_{0}\sigma^{a}\omega_{a}\right] \, . \label{cs27}%
\end{align}
The resulting supergravity Lagrangian is composed by three different sectors, each one controlled by the value of the coupling constants (coming from the S-expanded invariant tensors~(\ref{it4})), multiplied  by the level of the theory $\kappa$ which is related to Newton's constant $G$. The EH term appears in the first line of eq.(\ref{cs27}), and therefore $\nu_0$ can be normalized to one. In the critical point $\nu_{0}=\mu_{0}$ of the space of parameters, the theory decouples from fermions. This is a natural consequence of the form of the invariant tensor~(\ref{it3}). The Mielke-Baekler model~\cite{Mie91} is recovered at this critical point in the $\sigma \rightarrow 0$ limit.

\subsection{Field Equations}

Extremization of the action functional $S[A]=\int_{M_3}\mathcal{L}^{(3)}_{\mathrm{cs}}$ gives rise to the equations of motion. Alternatively, in the case of nondegeneracy, i.e., when all the coupling constants are nonvanishing and $\nu_{0}\neq\mu_{0}$, the field equations are more easily obtained through $F=\mathrm{d}A+A\wedge A=0$. Using eq.(\ref{cs1}) and (\ref{adsl1}), direct calculations shows%
\begin{align}
T^{a}-\frac{1}{2}\left(  \bar{\psi}\Gamma^{a}\psi\right)  +\epsilon_{~bc}^{a}\sigma^{b}e^{c}  &  =0\, ,\label{fe1}\\
R^{a}  &  =0\, ,\label{fe2}\\
\mathrm{D}\sigma^{a}-\frac{1}{2\ell}\left(  \bar{\psi}\Gamma^{a}\psi\right)  +\frac
{1}{2}\epsilon_{~bc}^{a}\left(  \sigma^{b}\sigma^{c}+\frac{1}{\ell^{2}}e^{b}e^{c}\right)
&  =0\, ,\label{fe3}\\
\mathrm{D}\bar{\psi}-\frac{1}{2}\left( \frac{1}{\ell}e^{a}+ \sigma^{a}\right)  \left(  \bar{\psi
}\Gamma_{a}\right)   &  =0 \, .\label{fe4}
\end{align}
Interestingly, the geometries characterized by the e.o.m are Lorentz flat. In addition, the presence of the Lorentz gauge fields $\sigma^{a}$ are source for the super-torsion $\hat{T}^{a}=T^{a}-\frac{1}{2}\left(  \bar{\psi}\Gamma^{a}\psi\right)$, as it can be seen from eq.(\ref{fe1}). In the limit $\sigma^{a}\rightarrow0$, solutions are described by constant torsion $T_{a}=\epsilon_{abc}e^{b}e^{c}$, flat Loretnz curvature $R^{a}=0$, and covariantly constant spinors $\nabla\bar{\psi}=\mathrm{D}\bar{\psi}-\frac{1}{2\ell} e^{a}   \left(  \bar{\psi
}\Gamma_{a}\right)=0 $. This is in contrast with the standard three-dimensional supergravity with AdS symmetry, in which solutions are Riemannian (torsionless) and constant Lorentz curvature.
\subsection{Symmetry transformations}

Under infinitesimal local gauge transformations, the one-form potential $A=A_{\mu}dx^{\mu}$ transforms
according to%
\begin{equation}
\delta A=\nabla\varrho=\mathrm{d}\varrho+\left[  A,\varrho\right]\, ,  \label{st1}%
\end{equation}
where
\begin{equation}
\varrho=\frac{1}{\ell}\rho^{a}P_{a}+\lambda^{a}J_{a}+\gamma
^{a}Z_{a}+\frac{1}{\sqrt{\ell}}\chi^{\alpha}Q_{\alpha}\label{st2}%
\end{equation}
is a zero-form gauge parameter. Using (\ref{adsl1}), (\ref{cs1}) and
(\ref{st2}) it is direct to show that
\begin{align}
\delta e^{a} &  =\mathrm{D}\rho^{a}-\epsilon_{~bc}^{a}\left(  \lambda^{b}+\gamma^{b}\right)  e^{c}-\epsilon_{~bc}^{a}\sigma^{b}\rho
^{c}+\left(  \bar{\psi}\Gamma^{a}\chi\right)  \, ,\label{st3}\\
\delta\omega^{a} &  =\mathrm{D}\lambda^{a}\, ,\label{st4}\\
\delta\sigma^{a} &  =\mathrm{D}\gamma^{a}+\epsilon_{~bc}^{a}\sigma^{b}\left(
\lambda^{c}+\gamma^{c}\right)  +\frac{1}{\ell^{2}}\epsilon_{~bc}^{a}e^{b}\rho
^{c}+\frac{1}{\ell}\left(  \bar{\psi}\Gamma^{a}\chi\right)\, ,  \label{st5}\\
\delta\bar{\psi} &  =\mathrm{D}\bar{\chi}-\frac{1}{2}\left(  \frac{1}{\ell}e^{a}+\sigma
^{a}\right)  \left(  \bar{\chi}\Gamma_{a}\right)  +\frac{1}{2}\left(  \frac
{1}{\ell}\rho^{a}+\lambda^{a}+\gamma^{a}\right)  \left(
\bar{\psi}\Gamma_{a}\right) \, , \label{st6}%
\end{align}
where $\mathrm{D}v^{a}=\mathrm{d}v^{a}+\epsilon_{~bc}^{a}\omega^{b}v^{c}$ for any zero-form parameter $v^{a}$ with one Lorentz index, and $\mathrm{D}\bar{\chi}=\mathrm{d}\bar{\chi}-\frac{1}{2}\omega^{a}\left(  \bar{\chi}\Gamma
_{a}\right)$ for any zero-form spinor $\chi^{\alpha}$. From eq.(\ref{st5}) is clear that the gauge field $\sigma^{a}$ transforms as a vector under local Lorentz rotations. Moreover, 
gauge transformations (\ref{st3})-(\ref{st6}) leave the CS Lagrangian (\ref{cs27}) invariant up to a closed form.

\newsection{Solutions and Killing Spinors}\label{sec6}

In supergravity, supersymmetric solutions are prescriptions for the bosonic fields of
the theory such that they solve the equations of motion arising from its action. In principle,
a solution need not be supersymmetric itself. A supersymmetric solution is defined as one which preserves a certain amount of the original supersymmetry. These are important when studying perturbative instabilities~\cite{Witten:1981me}. The standard calculation consists in embedding the bosonic theory into a supersymmetric one in such a way that the action remains stationary around a classical solution. In that case supersymmetry is, in general, enough to prove that this solution defines a local energy minimum and therefore is perturbatively stable.

We now focus on finding solutions to the Killing-Spinor equation for the AdS-Lorentz superalgebra in three-dimensions. In order to do so, we find a BTZ-type solution in the case of zero gravitino.  Moreover, we show that solutions around this classical configuration reduces to those found in~\cite{Coussaert:1993jp}. First we take $\bar{\psi}=0$ in (\ref{fe1})-(\ref{fe4})
%\begin{align*}
%T^{a}+\epsilon_{~bc}^{a}\sigma^{b}e^{c} &  =0\\
%R^{a} &  =0\\
%D\sigma^{a}+\frac{1}{2}\epsilon_{~bc}^{a}\left(  %\sigma^{b}\sigma^{c}+\frac
%{1}{\ell^{2}}e^{b}e^{c}\right)   &  =0
%\end{align*}
and solve for the ansatz
\begin{equation}
ds^{2}=-N^{2}dt^{2}+\frac{dr^{2}}{N^{2}}+r^{2}\left(  N_{\phi}dt+d\phi\right)
^{2} \label{btzans}
\end{equation}
with $\left(  x^{0},x^{1},x^{2}\right)  =\left(  t,r,\phi\right)  $,
$N=N(r)  $ and $N_{\phi}=N_{\phi}(r)$. We choose the following ansatz for the vielbein $e^{a}(x)$ and the spin connection $\omega^{a}(x)$
\begin{equation}
\begin{array}
[c]{ll}%
e^{0}=Ndt\, , &\qquad \omega^{0}=Vdt+Ydr+\beta V\,d\phi\, , \medskip\\
e^{1}=\frac{1}{N}dr\, , &\qquad \omega^{1}=Ddt+Wdr+\beta D\,d\phi\, , \medskip\\
e^{2}=r\left(  N_{\phi}dt+d\phi\right)\, ,   &\qquad \omega^{2}=Gdt+Hdr+\beta G\,d\phi\, ,
\end{array}
\label{vsc}
\end{equation}
where we have defined
\begin{equation}
\begin{array}
[c]{l}%
V=\sqrt{D^{2}+G^{2}-\beta_{1}} \, , \medskip\\
W=\frac{GG^{\prime}+DD^{\prime}}{VG}+\frac{DH}{G}\, , \medskip\\
Y=\frac{D^{\prime}+HV}{G}\, .%
\end{array}
\end{equation}
Here $D(r)$, $G(r)$ and $H(r)$ are arbitrary functions of the radial coordinate and $\beta$, $\beta_1$ are arbitrary constants~\cite{Hoseinzadeh:2014bla}.  Using the field equations (\ref{fe1})-(\ref{fe4}), in the case of zero gravitino $\bar{\psi}=0$, we have
\begin{equation}
\begin{array}
[c]{l}%
\sigma^{0}=-Vdt-Ydr+\left(  N-\beta V\right)  d\phi\, ,\medskip\\
\sigma^{1}=-Ddt-\left(  W+\frac{N_{\phi}}{N}\right)  dr-\beta Dd\phi\, ,\medskip\\
\sigma^{2}=\left(  \frac{r}{\ell^{2}}-G\right)  dt-Hdr+\left(  rN_{\phi
}-\beta G\right)  d\phi\, ,
\end{array}
\end{equation}
where
\begin{align}
N^{2}  & =-M+\frac{r^{2}}{\ell^{2}}+\frac{J^{2}}{4r^{2}} \, , \medskip\\
N_{\phi}  & =-\frac{J}{2r^{2}}\, , \label{sixsix}
\end{align}
being $M$ and $J$ are arbitrary real constants.

The Killing-spinor equation can be read from (\ref{st6}) when $\bar{\psi}=0$. This means
\begin{equation}
d\bar{\chi}-\frac{1}{2}\left(  \omega^{a}+\frac{1}{\ell}e^{a}+\sigma
^{a}\right)  \left(  \bar{\chi}\Gamma_{a}\right)  =0\, . \label{kse}
\end{equation}
Inserting (\ref{vsc})-(\ref{sixsix}) into (\ref{kse}) we find%
\begin{align}
\partial_{t}\chi^{\alpha}-\frac{\varepsilon}{2\ell}\chi^{\beta}\left[
N\Gamma_{0}+r\left(  \frac{1}{\ell}+N_{\phi}\right)  \Gamma_{2}\right]
_{\beta}^{~\alpha} &  =0\label{ks1} \, , \medskip\\
\partial_{r}\chi^{\alpha}+\frac{\varepsilon}{2}\chi^{\beta}\left[  \frac{1}%
{N}\left(  N_{\phi}-\frac{1}{\ell}\right)  \Gamma_{1}\right]  _{\beta
}^{~\alpha} &  =0\, , \medskip\label{ks2}\\
\partial_{\phi}\chi^{\alpha}-\frac{\varepsilon}{2}\chi^{\beta}\left[  r\left(
\frac{1}{\ell}+N_{\phi}\right)  \Gamma_{2}+N\Gamma_{0}\right]  _{\beta
}^{~\alpha} &  =0\, ,\label{ks3}%
\end{align}
where $\varepsilon=\pm1$, denotes the two nonequivalent gamma matrix
representation in three-dimensions. These equations are well known from~\cite{Coussaert:1993jp}. Following a similar analysis as presented in Ref.~\cite{Alvarez:2015bva}, solutions to (\ref{ks1})-(\ref{ks3}) are given by%
\begin{equation}
\bar{\chi}=\bar{\xi}\exp\left(  \vartheta^{+}\left[  \left(
-M+\frac{J}{\ell}\right)  \Gamma_{+\varepsilon}+\Gamma_{-\varepsilon}\right]
\Gamma_{0}\right)  \left(  z\Gamma_{+\varepsilon}+\frac{1}{z}\Gamma
_{-\varepsilon}\right) \label{kss}
\end{equation}
with
\begin{equation}
\begin{array}
[c]{ll}%
z=\left(  N+\frac{r}{\ell}+rN_{\phi}\right)  ^{\frac{1}{2}}\, , & \qquad \vartheta^{\pm
}=\frac{1}{2}\left(  \frac{t}{\ell}\pm\phi\right)\, , \medskip  \\
\Gamma_{\pm\varepsilon}=\frac{1}{2}\left(  1\pm\varepsilon\Gamma_{1}\right) \, ,
& \qquad \bar{\xi}=\text{constant spinor}\, .%
\end{array} \label{kssn}
\end{equation}

These expressions are only valid locally. In order to find globally defined killing spinor solutions, one needs to study their periodicity~\cite{Coussaert:1993jp}. In fact, under $\phi\rightarrow\phi+2\pi$, there is a phase $S=\left[  \left(
	-M+\frac{J}{\ell}\right)  \Gamma_{+\varepsilon}+\Gamma_{-\varepsilon}\right]
	\Gamma_{0}$ which gets multiplied by $2\pi$. Periodicity occurs if $S^{2}=0$, in which case there is no dependency on the angular coordinate $\phi$, for different values of $M$ and $J$. 
To conclude this section, we briefly discuss how exact supersymmetries for the zero-mass black hole $M=0=J$ are obtained. The extremal $M>0,\,M=|J|/\ell$, case can be treated similarly. In the zero-mass limit, the argument in the exponential of (\ref{kss}) is proportional to a nilpotent term $(\Gamma_{0}+\varepsilon\Gamma_{2})$. This implies $\exp\left(  \frac{\vartheta^{+}}{2}(\Gamma_{0}+\varepsilon\Gamma_{2})\right)=I+\frac{\vartheta^{+}}{2}(\Gamma_{0}+\varepsilon\Gamma_{2})$. Moreover, the linear dependency of $\bar{\chi}$ in $\vartheta^{+}$ drops out of the solution due to $\bar{\xi}$ is in the kernel of $(\Gamma_{0}+\varepsilon\Gamma_{2})$. In fact, $\bar{\xi}$ is an eigenvector of $\Gamma_1$ and has the form $\bar{\xi}=\binom{1}{\varepsilon}$. Finally, since $z=(\frac{2r}{\ell})^{\frac{1}{2}}$, we obtain 
\begin{equation}
\bar{\chi}=\left( \frac{2r}{\ell}\right) ^{\frac{1}{2}}\bar{\xi} \, .
\end{equation}
Therefore, there are two killing spinors, one for each value of $\varepsilon$. For the extremal black hole, periodicity is only reached when $\varepsilon=1$. This means that there is only one exact supersymmetry with $z=\left(\frac{2r}{\ell}-\frac{M\ell}{r}\right)^{\frac{1}{2}}$ and  $\bar{\xi}=\binom{1}{1}$. The killing spinors for the generic black hole solution (\ref{kss}) are non periodic or anti-periodic, hence there is no exact supersymmetries in that case. 
%\newpage 
\newsection{Discussion and future developments}\label{sec7}

In this work we have studied some aspects of the AdS-Lorentz superalgebra and the three-dimensional CS supergravity invariant under such symmetry.
Since the AdS-Lorentz superalgebra corresponds to the S-expansion of the $\mathcal{N}=1$ AdS superalgebra $\mathfrak{osp}\left(  2\left\vert 1\right.\right)  \otimes\mathfrak{sp}\left(  2\right)  $, it allowed us to show some features of the procedure in a simple but non-trivial case. 

Regarding the construction of the gauge supergravity theory, the essential ingredient is the invariant tensor. The S-expansion method provides the invariant tensors beyond the standard (super)trace using Ref.~\cite[Theorems 7.1 and 7.2]{Izaurieta:2006zz} and the Casimir operators following a treatment discussed in Ref.~\cite{Diaz2012}.
Using these invariants and the subspace separation method from Refs.~\cite{Iza05,Iza06a}, the CS three-form Lagrangian is written in terms of the Lorentz curvature and torsion. A new one-form gauge field has to be introduced as a consequence of having the $Z_{a}$ generator. This new Lorentz field $\sigma^{a}$ captures some of the features of the vielbein, namely, it transforms as a vector under representations of the Lorentz group and also interplay with the gravitino via supersymmetry transformations. From the dynamical point of view, the Lorentz gauge field $\sigma^{a}$ ia a source for the super-torsion $\hat{T}^{a}$ which makes  it propagate in vacuum.
Finally, an analytical stationary solution when the gravitino field is turned-off is discussed, and parallel Killing spinors around this background are computed. Solutions are shown to reduce the ones found in Ref.~\cite{Coussaert:1993jp}, which means that the supersymmetry properties of the BTZ-type balck hole presented in~\cite{Hoseinzadeh:2014bla}, are the same as the asymptotically AdS black holes of Einstein theory in three-dimensions.

We close with some brief discussion about possible applications of the AdS-Lorentz supergravity theory in three-dimensions. In the context of gauge/gravity dualities, in Ref.~\cite{Banados:1998pi} it has been shown that imposing suitable boundary conditions, the three-dimensional AdS supergravity lead to an asymptotic symmetry algebra given by two copies of the super-Virasoro algebra with central charge. This is a purely asymptotic phenomenon since the emergence of the conformal group at infinity is not the isometry group of any background geometry in three-dimensions. In the same context, the flat limit of AdS supergravity is discussed in~\cite{Barnich:2014cwa}, where a consistent set of asymptotic conditions are found. The asymptotic symmetry algebra in this case is given by the super-$\mathfrak{bms}_{3}$ algebra with a central extension. Interestingly, in Ref.~\cite{Caroca:2017onr} it has been shown that the $\mathfrak{bms}_{3}$ algebra can be obtained from the Virasoro by a S-expansion process. The analysis has been extended in such a way that new families of asymptotic symmetry algebras are constructed, all of them starting from the Virasoro one. As suspected, one of these new families corresponds to the asymptotic algebra of the CS theory for the Maxwell group~\cite{Concha:2018zeb}, whose boundary dynamics is described by an enlargement and deformation of the $\mathfrak{bms}_{3}$ algebra with three independent central charges. Since the relation between Maxwell and AdS-Lorentz algebra is well understood, it is reasonable to expect that the asymptotic symmetries for AdS-Lorentz CS theory can be obtained, following similar arguments as in~\cite{Concha:2018zeb}. This is work in progress and will be presented somewhere else. Finally, it would be interesting to extend previous discussion for studying the boundary dynamics of CS supergravity theories constructed using the supersymmetric extension of the Maxwell and AdS-Lorentz algebras.

\subsection*{Acknowledgments}
This work was supported in part by the Chilean FONDECYT Projects N\textsuperscript{o}\,3160437, 1150719, 1180681, and in part by the VRIIP-Unap Grant N\textsuperscript{o}\,039.428/2017. Also by UdeC through DIUC Grant N\textsuperscript{o}217.011. 056-1.0, and DINREG 19/2018 of the Direcci\'on de Investigaci\'on of the UCSC.
We thank P. K. Concha, J. D\'iaz, N. Merino, E. K. Rodr\'iguez, E. Rodr\'iguez and R. Troncoso for enlightening discussions and helpful comments.

\appendix
\newsection{Majorana spinors}\label{Appen}

The minimal irreducible spinor in three dimensions is a two real component
Majorana spinor. Every Majorana spinor satisfies a reality condition which can
be established by demanding that the Majorana conjugate equals the Dirac
conjugate
\begin{equation}
\bar{\psi}:= \psi^{\top}\mathcal{C}=-\mathrm{i}\,\psi^{\top}\Gamma_{1} \ .\label{maj1}%
\end{equation}
Spinors carry indices $\psi_{\alpha}$ and gamma-matrices act on them in such a
way that $\Gamma_{a}\psi:= \left(  \Gamma_{a}\right)  _{\beta}^{~\alpha
}\, \psi_{\alpha}$. In order to raise and lower indices, we introduce matrices
$(\mathcal{C}^{\alpha\beta})$, $(\mathcal{C}_{\alpha\beta})$ related to the
charge conjugation matrix, and we use the convention of raising and lowering
indices according to the NorthWest--SouthEast convention $\left(
\searrow\right)  $. This means that the position of the indices should appear
in that relative position as
\begin{equation}
\psi^{\alpha}=\mathcal{C}^{\alpha\beta}\, \psi_{\beta} \qquad
\mbox{and} \qquad \psi_{\alpha}%
=\psi^{\beta}\, \mathcal{C}_{\alpha\beta} \ , \label{maj2}
\end{equation}
which implies that
\begin{equation}
\mathcal{C}^{\alpha\beta}\, \mathcal{C}_{\gamma\beta}=\delta_{\gamma}^{\alpha
} \qquad \mbox{and} \qquad \text{ }\mathcal{C}_{\beta\alpha}\, \mathcal{C}^{\beta\gamma}=\delta_{\alpha
}^{\gamma} \ .
\end{equation}
We choose the identifications in such a way that the Majorana conjugate
$\bar{\psi}$ is written as $\psi^{\alpha}$. Comparing eq.~(\ref{maj1})
with eq.~(\ref{maj2}), one then finds
$(\mathcal{C}^{\alpha\beta})=\mathcal{C}^{\top}$ and $(\mathcal{C}_{\alpha\beta
})=\mathcal{C}^{-1}$.

\bibliographystyle{utphys}
\bibliography{bibSS}

\end{document}